# Ostwald ripening in an oxide-on-metal system

Natalia Michalak,[1,2] Tomasz Ossowski,[3] Zygmunt Miłosz,[1] Mauricio J. Prieto,[4] Ying Wang,[1] Mirosław Werwiński,[2] Visnja Babacic,[1] Francesca Genuzio,[5] Luca Vattuone,[6,7] Adam Kiejna,[3] Thomas Schmidt,[4] and Mikołaj Lewandowski[1,*]

[1]NanoBioMedical Centre, Adam Mickiewicz University, Wszechnicy Piastowskiej 3, 61-614 Poznań, Poland

[2]Institute of Molecular Physics, Polish Academy of Sciences, M. Smoluchowskiego 17, 60-179 Poznań, Poland

[3]Institute of Experimental Physics, University of Wrocław, Pl. M. Borna 9, 50-204 Wrocław, Poland

[4]Department of Interface Science, Fritz-Haber-Institut der Max-Planck-Gesellschaft, Faradayweg 4–6, 14195 Berlin, Germany

[5]Elettra–Sincrotrone Trieste S.C.p.A., S.S. 14 - km 163, 5 in AREA Science Park, 34149 Basovizza, Trieste, Italy

[6]Dipartimento di Fisica, Università di Genova, Via Dodecaneso 33, 16146 Genova, Italy

[7]IMEM-CNR, Via Dodecaneso 33, 16146 Genova, Italy

[*]Corresponding author: lewandowski@amu.edu.pl

ORCID: 0000-0002-0984-3456 (N.M.), 0000-0002-2301-8671 (T.O.), 0000-0002-4938-9331 (Z.M.), 0000-0002-5087-4545 (M.J.P.), 0000-0002-8905-0818 (Y.W.), 0000-0003-1934-4818 (M.W.), 0000-0002-6986-550X (V.B.), 0000-0003-0699-2525 (F.G.), 0000-0003-3718-6470 (L.V.), 0000-0002-0983-7953 (A.K.), 0000-0003-4389-2080 (T.S.), 0000-0002-1480-8516 (M.L.)

## ABSTRACT

Ostwald ripening is a well-known physicochemical phenomenon in which smaller particles, characterized by high surface energy, dissolve and feed the bigger ones that are thermodynamically more stable. The effect is commonly observed in solid and liquid solutions, as well as in systems consisting of supported metal clusters or liquid droplets. Here, we provide the first evidence for the occurrence of Ostwald ripening in an oxide-on-metal system which, in our case, consists of ultrathin iron monoxide (FeO) islands grown on Ru(0001) single-crystal support. The results reveal that the thermally-driven sintering of islands allows altering their fine structural characteristics, including size, perimeter length, defect density and stoichiometry, which are crucial, e.g., from the point of view of heterogeneous catalysis.

The understanding of physicochemical phenomena that determine the size, structure and properties of nanomaterials is crucial for their rational design. One of the interesting groups of nanomaterials are the 2D transition metal oxides, such as ultrathin islands and films grown on metal substrates [1,2]. Thanks to their low-dimensionality and the interaction with the underlying support, such systems often exhibit unique physicochemical properties that predispose them for applications in nanoelectronics/spintronics, energy storage/conversion and heterogeneous catalysis. Low-dimensional metal-supported oxide species may also constitute well-defined model systems for performing fundamental studies on "inverse" catalysts [3], as they resemble many structural features of real catalysts, such as oxide species of different size and stoichiometry, low-coordinated transition metal atoms (present at the perimeter of oxide islands and within internal defects), as well as exposed noble metal support regions.

Depending on the application, it may be desirable to maximize the number of specific structural features within an oxide-on-metal system, such as certain crystal facets, edge sites, defects or exposed substrate regions [4]. This is usually done by optimizing the oxide's growth method. However, the structure of the system may also evolve under varying environmental conditions, e.g. in a response to high pressures of gases or elevated temperatures. With this respect, controlling the structure of metal-on-metal and metal-on-oxide systems, and predicting their behavior in different environments, is much more trivial, as the mechanisms governing the growth and structural evolution of metals under oxidizing and reducing conditions are well-understood [5]. For example, thermal treatments under reducing conditions are known to lead to supported metal particles sintering and, thus, the change in the ratio between the facet and edge sites. The process usually proceeds through one of the two mechanisms: coalescence [5], i.e. Brownian-like motion of particles and their merging upon collision, or Ostwald ripening [5,6], in which smaller particles dissolve and feed the bigger ones. This behavior is related to a much higher surface energy of smaller particles and higher thermodynamic stability of the bigger ones. The influence of oxidizing treatments is also well documented, leading to the formation of surface and bulk oxides. In the case of metal-supported oxide species, the situation is much more complex, as the structural transformations critically depend on the physicochemical properties of a specific oxide (e.g. redox potential of constituent metal elements, as well as the interaction of the oxide with the metal support on which it is grown). It is, thus, crucial to carry out fundamental studies aimed at identifying the possible transformations occurring in oxide-on-metal systems under oxidizing and reducing conditions, and elucidating the corresponding physicochemical mechanisms.

In this Letter, on the example of ultrathin FeO islands grown on Ru(0001) [7,8], we show the Ostwald ripening can also occur in an oxide-on-metal system. The islands sinter in response to high temperature annealing in ultra-high vacuum (UHV), leading to the change in their average size and total perimeter length, as well as the number of internal defects and the exposed support area. The process is accompanied by the transformation of FeO into an O-poor $Fe_xO$ $(1 < x \leq 2)$ phase, which can be potentially reoxidized back to FeO while preserving the newly-established morphology.

The experiments were performed in two different UHV chambers (base pressures in the $10^{-10}$ mbar range) equipped with sample preparation (Ar sputter guns, e-beam heating stages, Fe evaporators, $O_2$ lines) and characterization facilities: scanning tunneling microscopy/spectroscopy (STM/STS) and low energy electron diffraction (LEED) in one chamber and low energy electron microscopy/X-ray photoemission electron microscopy (LEEM/XPEEM), as well as X-ray photoelectron spectroscopy (XPS) with synchrotron radiation in the other (UE49 SMART beamline at BESSY II [9,10]). Ru(0001) single crystal was cleaned by repeated cycles of $Ar^+$ sputtering, annealing in $O_2$ and in UHV. Ultrathin FeO islands were grown by Fe deposition at room temperature and post-oxidation. The islands were annealed in UHV, with the temperature being monitored using infrared pyrometers. The structural changes were studied in real time with LEEM and using STM/STS, LEED, XPEEM and XPS.

Theoretical calculations on the structure and energetics of FeO and $Fe_2O$ films were performed in a 3×3 computational cell using spin-dependent density functional theory (DFT) and projector-augmented-wave (PAW) method [11,12], as implemented in the Vienna ab initio simulation package (VASP) [13–15]. The Perdew-Burke-Ernzerhof (PBE) form [16] of the generalized gradient approximation (GGA) functional and the Hubbard U correction, were used [17]. The Fe 3p electronic states, magnetic moments and Mulliken charges on atoms were calculated using the full-potential local-orbital scheme code (FPLO, ver. 18.00-52) [18]. The relaxed geometries obtained from VASP constituted an input for FPLO, while the use of an all-electron code allowed going beyond the limitations of VASP in describing the electronic core-level states. Detailed description of experimental and computational methods and procedures is provided in the Supplemental Material file.

Figure 1(a) shows a STM image of pristine FeO islands grown on Ru(0001) by Fe deposition and post-oxidation in $1 \times 10^{-6}$ mbar $O_2$ at 950 K, while Figures 1(b) and 1(c) show the system after annealing at 950 K in UHV for 10 and 20 min, respectively. Pristine islands are well-dispersed, having different sizes and growing both at the regular terrace sites and at the step edges of the substrate [7,8]. They have a height of ~5 Å, indicating they are composed of two FeO layers, and exhibit a Moiré superstructure with 21.6 Å periodicity (originating from the lattice mismatch between the oxide ($a_{FeO(111)}$=3.04 Å) and the Ru(0001) support (2.71 Å). The observed growth mode and structure are characteristic for the preparation conditions used [7,8]. Bigger islands are formed through merging of smaller ones, which results in the appearance of characteristic line defects within the islands (Figure 1(e)) [7,8,19]. The distribution of defects may be traced using STS dI/dV mapping mode with a bias voltage value of +1.0 V (Figure 1(f)). Following annealing at 950 K in UHV for 10 min, the number of islands decreases, while the average island size increases, with the total coverage staying the same (within an accuracy of 5%). Moreover, the island edges become sharper, running along the crystallographic directions of the ruthenium support. Annealing for additional 10 min at the same temperature leads to further growth of bigger islands at the expense of smaller ones and the visible change in the islands shape towards the equilibrium one, i.e. truncated triangle (or – ultimately – hexagonal). A significant decrease of the number of internal structural defects is also observed, ~80% of which vanish after the first annealing and almost none are left after the second one (Figure 1(g)). The total edge length decreased by approx. 20% after the first annealing and stayed more or less the same after the second one. Most importantly, the height profiles taken across both structures (Figure S1 in the Supplemental Material file) reveal that the annealed islands are by ~0.5 Å lower than the pristine ones (4.45+/-0.15 Å (annealed) vs. 5.0 Å (pristine)). This indicated that sintering is accompanied by a transition to a structure with lower apparent height. The structural evolution takes place gradually, as evidenced by the histogram shown in Figure 1(d). Notably, the annealed islands were characterized by a virtually-identical Moiré structure as pristine FeO (compare Figures 1(e) and 1(g)), indicating that the new phase is structurally similar to FeO.

The LEED pattern obtained for a pristine sample, shown in Figure 1(h), reveals the presence of Ru(0001)-(1×1) and FeO(111)-(1×1) spots, the "satellite" reflexes originating from the presence of the FeO Moiré superstructure [7] and the (2×2) spots corresponding to O atoms chemisorbed on the exposed substrate regions and forming (2×1) or (2×2) domains [20]. The diffraction pattern of the annealed sample (Figure 1(i)) exhibits reflexes located at similar positions, which indicates that there are no significant changes in the atomic or Moiré periodicities of iron oxide. Thus, the results further confirmed the structural similarity of pristine and annealed islands. The only visible difference is the increased intensity of the spots related to the formation of larger ordered iron oxide regions.

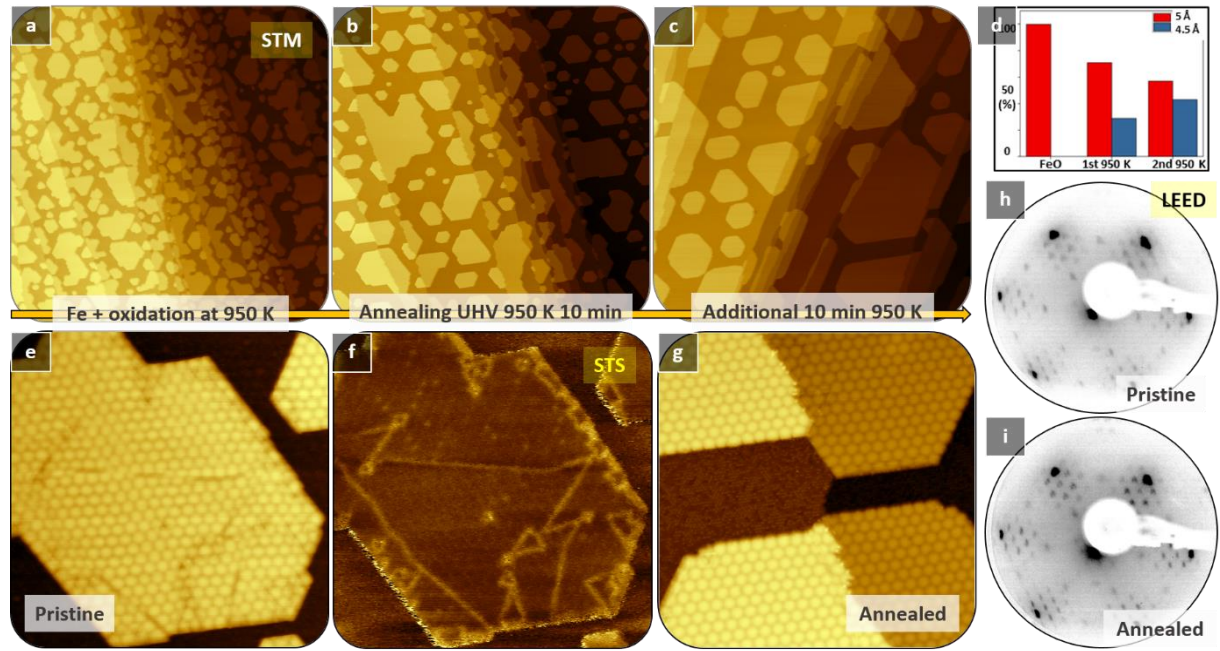

**Figure 1.** STM images of pristine FeO/Ru(0001) (a) and the system after annealing at 950 K for 10 (b) and 20 min (c) in UHV. (d) presents the island height histogram. (e) and (f) show the structural defects within an extended island observed in STM and STS dI/dV mapping modes, respectively. (g) reveals the lack of defects following annealing at 950 K for 10 min. (h) and (i) present LEED patterns obtained for pristine and annealed samples, respectively. STM: $V_{sample}$ = +1.0 V; $I_t$ = 0.7 nA (all images, including STS dI/dV maps); Image sizes: 500×500 $nm^2$ (a-c), 50×50 $nm^2$ (e-g). LEED: 64 eV.

The observed sintering of the islands could be explained either by coalescence or by Ostwald ripening. To clarify this issue, we used LEEM which allows real-time monitoring of structural changes that occur at solid surfaces during thermal or chemical treatments. Figure 2(a) shows the pristine FeO/Ru(0001), while Figures 2(b) and 2(c) the system after UHV annealing at 1050 K for 45 seconds and 2 minutes, respectively. The movie showing the transformation (at adjusted speed) can be found in the online Supplemental Material section. The higher temperature used in LEEM studies was dictated by the slow dynamics of the process at 950 K. During the annealing, the smallest islands (marked with yellow and blue circles in Figures 2(a) and Figure 2(b), respectively) disappear and, after some time, the bigger ones start to grow laterally in size (Figures 2(d-f)). This kind of behavior is characteristic for the Ostwald ripening process. The diffusion of the decomposed material across the surface was not visible due to instrumentation limits. The growing parts of the islands show different contrast in LEEM when imaged with certain instrument parameters (Figure 2(d)), which confirms that they are characterized by a structure that is different than that of original islands. During a prolonged annealing, the contrast within the islands equalizes towards that of the newly-grown part (Figures 2(h-i)), which is in agreement with the gradual nature of the process observed with STM (Figure 1(d)).

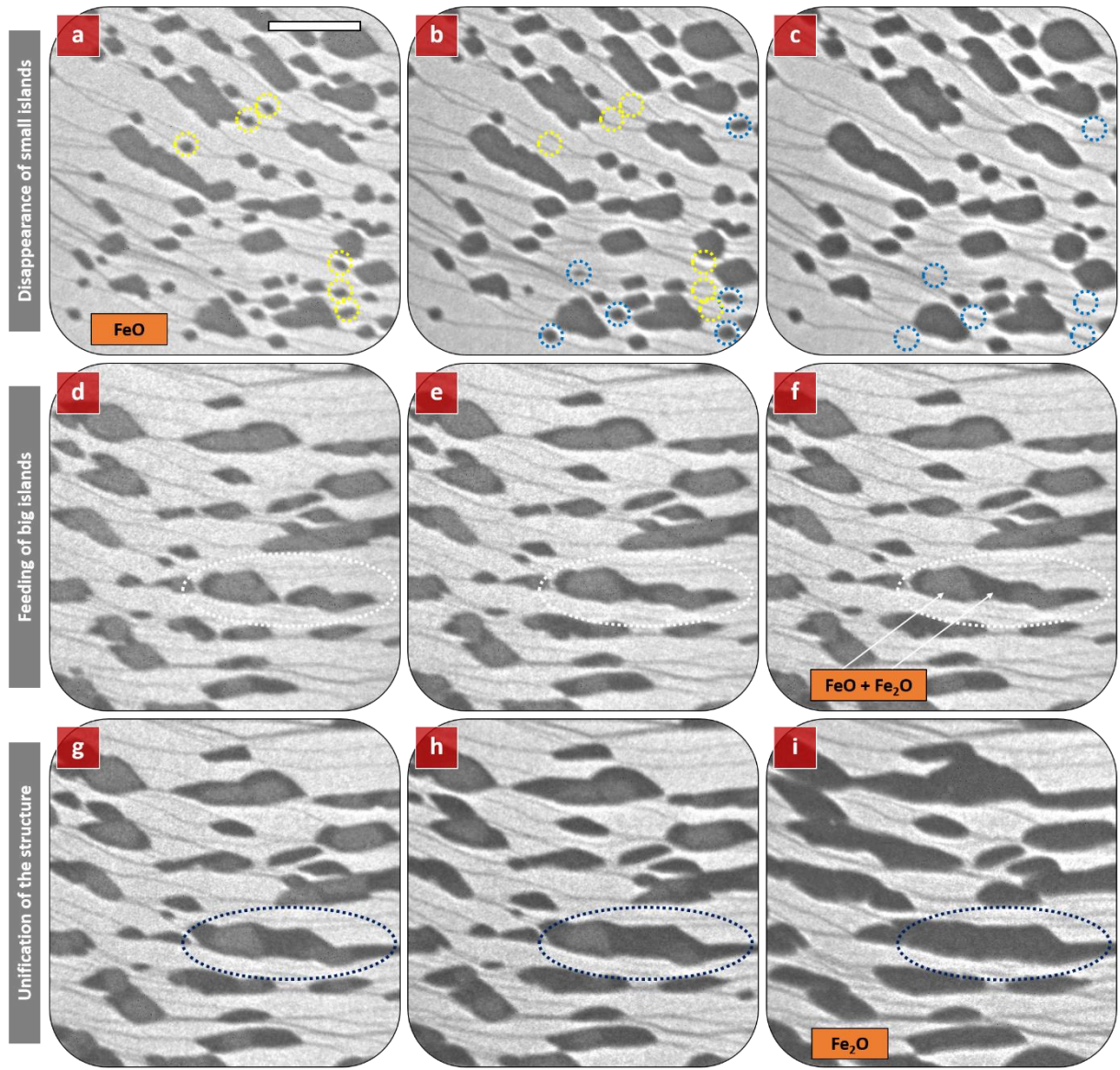

**Figure 2.** LEEM images obtained for (a) pristine FeO/Ru(0001) and (d-i) the system at different time intervals of annealing at 1050 K. Due to thermal drift, the field of view shifted to the left between (c) and (d) (the islands seen on the right-hand-side in (c) are visible on the left-hand-side in (d)). The observed structural transformation can be divided into three phases: (a-c) disappearance of smaller FeO islands, (d-f) growth of bigger islands, with the new parts exhibiting different contrast, and (g-i) unification of the structure towards the newly-grown parts. Scale bar: 500 nm. LEEM energy: 20 eV.

To determine the difference in the structure of pristine FeO and the phase formed during the annealing, synchrotron radiation was utilized for recording XPEEM images of a partially transformed sample (similar state to the one shown in Figure 2(d)). A series of images corresponding to the energy range in which the Fe 3p photoelectron signals are expected to appear was collected and the signal-vs-intensity curves were locally plotted from the regions corresponding to the original and growing parts of an exemplary island (Figure 3). The results indicate that the new phase hosts iron in a lower oxidation state than the original one (the edge of the curve is shifted towards higher binding energy values). The maximum of the signal recorded for the original structure, ~54.5 eV, lies in between the ones expected for $Fe^{2+}$ (53.7 eV) and $Fe^{3+}$ (55.6 eV) [21]. This is in agreement with our recent studies which indicated mixed valency of iron ($Fe^{2+}/Fe^{3+}$) in bilayer FeO on Ru(0001) [19]. The signal recorded from the growing parts, on the other hand, indicates the dominant contribution from iron in the $Fe^{2+}$ oxidation state. A similar trend is observed in the recorded XPS Fe 3p (Figure S2(a)) and Fe 2p (Figure S2(b)) spectra, which show a gradual shift of both signals to lower binding energies following brief and prolonged annealing. Moreover, the Fe 2p data correlated with the O 1s signals (not shown) allowed determining the change in the Fe:O signal ratio before and after the complete structural transformation. For this purpose, the total area under the signals was integrated and divided by the atomic sensitivity factors of the respective photoelectrons. Taking into account the initial and final FeO coverage (with their local fluctuations on different sample regions), as well as the fact that the remaining part of the surface is covered by chemisorbed oxygen forming (2×1) domains (and probably also less densely-packed (2×2) after annealing), the stoichiometry of the new phase was determined to be roughly "$Fe_2O$". The precise stoichiometry could not be established due to potential partial diffusion of iron and oxygen into the substrate, local changes of oxygen concentration on Ru(0001) (partial transformation of the (2×1) structure into the (2×2)) and the possibility that the transformation was not fully completed for all the islands. Thus, the results revealed that the islands formed following a thermally-driven sintering represent an O-poor $Fe_xO$ ($1 < x \leq 2$) phase.

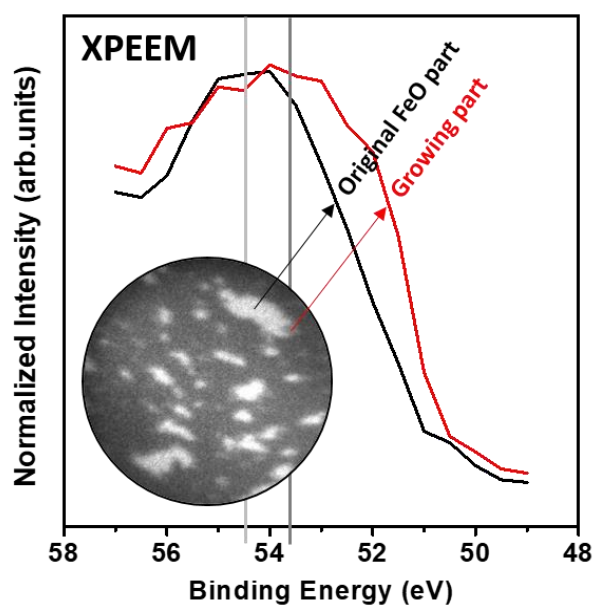

**Figure 3.** XPEEM-IV Fe 3p spectra obtained for different regions of iron oxide island formed by annealing FeO/Ru(0001) at ~1050 K in UHV. The black curve was taken from the part where the original FeO island was present, while the red one from the newly-grown part.

In order to get more insight into the structure of the newly-formed iron oxide phase, supporting DFT calculations were performed. The information obtained from experimental studies, i.e. atomic and superstructure periodicities similar to pristine FeO(111) (deduced from STM and LEED results), the ~0.5 Å lower height (determined with STM), the lower oxidation state of iron (from XPEEM and XPS) and the Fe:O ratio (from XPS), allowed us to limit the number of investigated systems to three differently stacked layers with "Fe$_2$O" stoichiometry: Fe-O-Fe/Ru, Fe-Fe-O/Ru and O-Fe-Fe/Ru. The large period of the FeO/Ru(0001) Moiré superstructure and the necessity to include three atomic layers forming the oxide (two layers of Fe and one layer of O), as well as five layers of the Ru substrate (with the bottom three fixed in bulk positions and two allowed to relax), made the size of the computational cell challenging. To make computations feasible, separate calculations were carried out for different high-symmetry regions (HSRs) of the Moiré supercell, i.e. *top*, *fcc* and *hcp*. Such approach was found to reproduce the crucial structural features of FeO on different single-crystal substrate: Pt(111) [22]. Here, we discuss the results obtained for the *fcc* HSR (which was found to be the most energetically-stable one) and the laterally iron oxide compressed to the Ru(0001) lattice constant ($a_{Ru(0001)}$=2.73 Å – as determined from preliminary calculations). The results obtained for other HSRs are presented in Figure S3. The analysis of the stability shows that the Fe-O-Fe/Ru and O-Fe-Fe/Ru stackings are the energetically most preferred ones, being very close in total energy values (with the Fe-O-Fe/Ru as the most preferred one by 0.35 eV (Figure 4(a)). The Fe-Fe-O/Ru system with oxygen atoms at the interface is energetically unfavored (the total energy is by 13.11 eV higher than that of the most stable one). The calculations also reveal that the atomic planes in the relaxed structures remain uncorrugated within a single HSR. The estimated layer thickness for the Fe-O-Fe/Ru configuration is 5.05 Å (Figure S3), which is only ~15% lower than that obtained for bilayer FeO (6.08 Å [23]). The calculations performed using the lattice constant of FeO (substrate atoms laterally expanded to 3.04 Å) resulted in a similar height difference (independently of the Moiré HSR; Figure S4). Even though the experimentally determined height difference is only 10%, the Fe-O-Fe/Ru stacking of the Fe$_2$O layers seems reasonable, as the height measured with STM may be altered by electronic effects, especially when measuring the height between the oxide and the metal. Notably, the order of the first three layers (looking from the substrate) is the same for the most energetically-stable FeO and Fe$_2$O configurations, with FeO having an additional O layer on top.

Since the Fe-O-Fe/Ru and O-Fe-Fe/Ru stackings are almost degenerate in energy, additional studies, aimed at their distinguishment, were performed. The calculated work function values of both structures were found to be 4.65 and 7.65 eV, respectively (as obtained for the *fcc* HSR; for other HSRs, see Table S1 in the SM), which is related to their different surface dipole. The value obtained for the most stable FeO stacking (O-Fe-O-Fe/Ru), on the other hand, is ~7.8 eV. Thus, the difference between the work function of FeO and the new phase can be a guide for the Fe$_2$O stacking determination. From the recorded LEEM-IV curves (Figure S5), it is evident that the work function of the newly-grown phase is by about 0.95 eV lower than the one of pristine FeO (in LEEM-IV, the work function difference can be determined from the mutual position of characteristic signal intensity drops observed at low energies). All this suggests that even though the Fe-O-Fe/Ru and O-Fe-Fe/Ru are very close in total energy, their calculated and measured work function values favor the former as the one observed experimentally.

To investigate the influence of the FeO-to-Fe$_2$O transformation on the shape of XPS Fe 3p line, the core-levels were calculated using an all-electron code [24] and the density of states (DOS) of Fe 3p orbitals of both structures was determined. The results were correlated with Fe 3p XPEEM and XPS spectra obtained experimentally. As can be seen from Figures 4(b) and S6, the Fe 3p states of the Fe$_2$O phase are more widely spaced and shifted towards lower binding energies compared to those of FeO, which is in agreement with the experimental observations. The Fe 3p spectra of both structures consist of superimposed contributions originating from iron atoms in both, i.e. top and bottom, layers. Each Fe site provides six 3p states, the configuration of which depends, inter alia, on spin-orbit coupling and spin polarization (see Figure S7) [24–26]. Thus, the chemical shift of 3p states results from charge redistribution caused by the rearrangement of chemical bonds (reduction of iron oxide) [24].

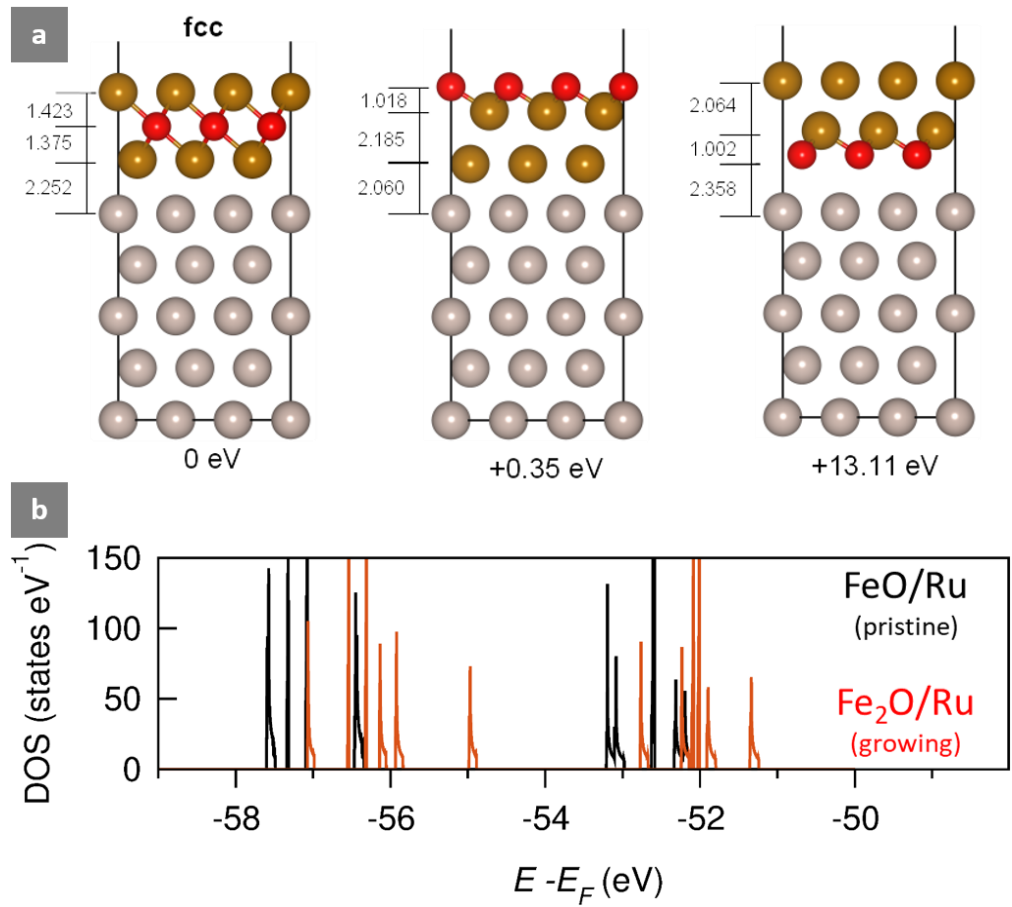

**Figure 4.** (a) Relaxed geometries obtained from DFT calculations for iron oxide structures with $Fe_2O$ stoichiometry on Ru(0001). Different stacking are considered: Fe-O-Fe/Ru (left), O-Fe-Fe/Ru (middle) and Fe-Fe-O/Ru (right). Gold spheres represent Fe, red spheres O and silver spheres Ru atoms, respectively. The energies are referenced to the most stable system. The distances are given in Å. (b) Calculated DOS of Fe 3p orbitals of FeO and $Fe_2O$ structures on Ru(0001) (the calculations were performed using the FLPO code with U(Fe 3d) = 4 eV; the results obtained for different U values are presented in Figure S6).

The calculated magnetic moments and Mulliken charges on atoms, presented in Table S2, reveal that the most significant changes are observed for the outer (top) iron layer. The decrease in the value of the spin magnetic moment from -4.21 to -3.55 $\mu_B$ on $Fe_{top}$ atoms causes a narrowing of the 3p spectrum. On the other hand, the decrease in the positive charge on the atoms (from +0.96 to +0.39) leads to the shift of the spectrum towards lower binding energies, which provides a route for the widening of the total Fe 3p signal ($Fe_{top}+Fe_{bottom}$). The widening and shift in the Fe 3p spectra between $Fe_2O$ and FeO is, thus, a consequence of the changes in magnetic moments and charges on Fe atoms in the top FeO layer.

The origin of the Ostwald ripening process and the FeO-to-Fe$_2$O transformation is believed to be related to energy minimization, i.e. preferred reduction of the number of FeO/Ru edge sites, as well as the structural characteristics of the FeO/Ru(0001) system. First of all, the oxide on this particular substrate preferably grows in the bilayer (O-Fe-O-Fe) form, and not as a monolayer (O-Fe) like on most single-crystal supports (e.g. Pt(111) [27]). The monolayer can be stabilized, however, only within a limited range of oxygen pressures [8]. Secondly, the top FeO layer hosts Fe$^{3+}$ ions, [19] which have a much higher standard reduction potential to Fe$^{2+}$ than the Fe$^{2+}$ to metallic Fe. This opens a relatively simple route for the thermal reduction of the top FeO layer. Following such reduction, the surface free energy of the island would change due to the modification of the surface termination [29]. Finally, the oxide is polar when grown in the [111] direction [7], so any charge redistribution within the system may lead to significant restructuring (due to increased contribution of electrostatic energy to surface energy [28]). Thus, the initial reduction of the top layer could start a kind of "chain reaction" which leads to the complete structural decomposition of an oxide island. The resulting material (most probably metallic iron) would then diffuse across the surface, binding to the encountered still-existing larger islands that are thermodynamically more stable. It could then become oxidized by O atoms present on Ru(0001), which constitute an oxygen "reservoir". The final structure partially results from the vicinity of pristine FeO with its original layers stacking, i.e. O-Fe-O-Fe/Ru [7], and atomic periodicity (the growing part would try to adopt the stacking and follow the atomic period of the neighboring FeO), as well as the limited amount of available oxygen would disallow the full reconstruction of the FeO structure [8], which would lead to the formation of the Fe$_2$O phase.

It is important to mention, that the observed sintering mechanism – being the classical Ostwald ripening process – is different than the one observed by the Authors of Ref. [30] for VO$_x$ islands on Rh(111). In their case, sintering was only observed under specific reaction conditions, with the islands initially moving towards each other and ultimately coalescing via the so-called polymerization-depolymerization mechanism. Thus, the process is driven by a chemical reaction and not a simple evaporation-condensation mechanism like in our case.

In summary, we have shown that UHV annealing leads to sintering of bilayer FeO islands grown on Ru(0001), which affects the general structural characteristics of the system: the average island size, perimeter length, number of internal defect and the stoichiometry of the oxide phase. The process occurs via the Ostwald ripening mechanism (the smaller islands decompose, the material diffuses across the surface and feeds the encountered larger ones), which was not observed for an oxide-on-metal system so far. Even though the annealed islands represent the "Fe$_2$O" phase, they could be potentially reoxidized after annealing, thus regaining the FeO stoichiometry, however, without becoming dispersed again. The work constitutes an important step towards understanding of complex physicochemical processes occurring in oxide-on-metal systems under varying environmental conditions.

## Acknowledgements

The studies were financially supported by the National Science Centre of Poland (from 2015 to 2017 through the SONATA 3 project No. 2012/05/D/ST3/02855 – granted to M.L. – and from 2017 to 2019 through the PRELUDIUM 11 project No. 2016/21/N/ST4/00302 – granted to N.M.), as well as the Foundation for Polish Science (from 2019, through the First

TEAM/2016-2/14 (POIR.04.04.00-00-28CE/16-00) project "Multifunctional ultrathin Fe(x)O(y), Fe(x)S(y) and Fe(x)N(y) films with unique electronic, catalytic and magnetic properties" co-financed by the European Union under the European Regional Development Fund – granted to M.L.). The authors thank the Helmholtz-Zentrum-Berlin for the allocation of a synchrotron radiation beamtime, as well as Gina Peschel and Feng Xiong for their assistance during the measurements. T.O. and A.K. acknowledge the computer time granted by the ICM of the Warsaw University (projects GB77-16 and GB82-9). The calculations performed by M.W. were partially carried out using the computational resources provided by the Poznań Supercomputing and Networking Center (PSNC).

**Supplemental Material**

Movie (published online) showing the Ostwald ripening process in real time, detailed description of experimental and computational methods and procedures, additional STM images with height profiles taken over FeO and $Fe_2O$ islands, XPS Fe 3p and 2p spectra, differently-stacked $Fe_2O$ structures obtained from DFT calculations, LEEM-IV curves, calculated DOS of Fe 3p orbitals in FeO/Ru and $Fe_2O$/Ru structures, work function values, magnetic moments and Mulliken charges.

**GRAPHICAL ABSTRACT**

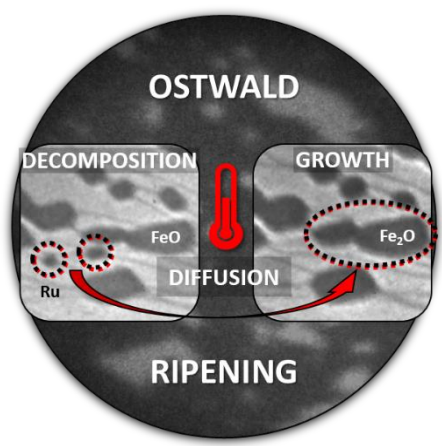

## Supplemental Material

**Detailed description of experimental and computational methods and procedures**

The experiments were performed in two different UHV chambers with base pressures in the $10^{-10}$ mbar range. Both chambers were equipped with *in situ* sample preparation (Ar sputter guns, e-beam heating stages, Fe evaporators, $O_2$ lines) and characterization facilities. One chamber (located in Poznań, Poland) was hosting STM and LEED (both from Omicron), while the other (located at the BESSY II synchrotron light source in Berlin, Germany) was equipped with a LEEM instrument allowing, in addition to basic imaging, recording μLEED patterns, as well as XPEEM images and XPS spectra using synchrotron light (beamline UE49 SMART). The Ru(0001) single crystal (purity ≥ 99.99%; from MaTeck) was cleaned by repeated cycles of $Ar^+$ (99.999%; Messer) ions sputtering followed by annealing in $O_2$ at 1000 K and at 1450 K in UHV. Fe was deposited from a 2 mm rod (99.999%; Alfa Aesar) using an electron beam evaporator and oxidized in $1\times10^{-6}$ mbar $O_2$ (>99.995%, Linde gas). The cleanliness of the crystals and the formation of well-ordered iron oxide films were confirmed by STM and LEED (in Poznań) or LEEM and μLEED (in Berlin). The samples were annealed in UHV at 950 (Poznań) and 1050 K (Berlin), with the temperature monitored using infrared pyrometers (LumaSense). The STM images were recorded at room temperature, in constant current mode, using electrochemically etched W tips. The images were processed with the Gwydion computer software. The STS curves were collected using the lock-in technique (oscillation frequency: 6777 Hz, amplitude: 40 mV). XPEEM and XPS curves were recorded using photon energies of 120 (for recording Fe 3p spectra), 600 (O 1s) and 780 (Fe 2p) eV. The spectra were calibrated based on the position of the valence band, by taking into account the parameters of the spectrometer and the mutual position of the signals. The XPS results were analyzed using the CasaXPS software (Casa Software Ltd.). Shirley background subtraction and Gaussian-Lorentzian lineshape were used for the analysis. For presentation, all XPS, XPEEM-IV and LEEM-IV spectra were normalized, with each XPEEM-IV curve being averaged from four recorded curves.

The theoretical calculations of the structure of FeO and $Fe_2O$ films on Ru(0001) were performed using spin-dependent DFT, as implemented in the Vienna ab initio simulation package (VASP) [1–3]. The electron ion-core interactions were represented by the projector-augmented-wave (PAW) potentials [4,5]. A plane wave basis set with a kinetic energy cutoff of 500 eV was applied. The exchange-correlation energy was treated at the spin-polarized generalized gradient approximation (GGA) level using the Perdew-Burke-Ernzerhof (PBE) functional [6]. The realistic description of the electronic structure of FeO requires consideration of strong correlations of Fe 3d electrons. This was realized by the Hubbard U correction within the rotationally invariant approach of Dudarev et al. [7], with the effective parameter $U_{eff} = 4$ eV. The Brillouin zone was sampled using Γ-centered k-point meshes of $24\times24\times16$ for bulk hcp Ru calculations. A Fermi surface broadening of 0.2 eV was applied to improve convergence of the solutions using the second order Methfessel-Paxton method [8]. The calculated lattice constant of rock-salt FeO, 4.35 Å, is in agreement with other theoretical calculations (4.35 Å [9]), as well as experimental observations (4.35 Å [10]). Analysis of the magnetic structure shows that the oxide is antiferromagnetic in the [111] direction, with a magnetic moment per Fe atom of 3.69 $\mu_B$ (which fits reasonably well the value determined experimentally at 4.2 K and reported in Ref. [11], i.e. 3.32 $\mu_B$). The calculated lattice parameters of bulk hcp ruthenium, a = 2.73 Å and c/a = 1.58, also agree with the published experimental data (a = 2.71 Å and c/a = 1.58) [12]. Ru(0001) surface was modelled by an asymmetric slab consisting of five atomic layers and a 3×3 surface unit cell. The two bottom layers were frozen in their bulk positions, while the atomic positions of the remaining atoms were optimized until the residual Hellman-Feynman forces on atoms were smaller than 0.01 eV/Å. Iron oxide layers were placed on top of the substrate in *fcc*, *top* and *hcp* geometries, which aimed in reproducing the three high-symmetry regions of the FeO/Ru(0001) Moiré supercell. The 3×3 surface unit cell is a good approximation, however, it requires to fit one geometry (FeO) to the other (Ru). From the calculations, the interatomic distances within the iron and oxygen planes in FeO(111) equal to about 3.07 Å and in Ru(0001) to about 2.73 Å, were obtained. Thus, adjusting the structures requires either to compress the oxide or expand the substrate by about 10%. In order to get reliable results, two sets of calculations were performed: with the FeO (or Fe₂O) lattice constant compressed to the one of Ru(0001) and with the Ru(0001) lattice constant expanded the meet the one of FeO. Calculations of Fe 3p electronic states were performed using a full-potential local-orbital scheme code (FPLO, ver. 18.00-52) [13]. The relaxed geometries obtained from VASP constituted an input for FPLO, while the use of an all-electron code allowed going beyond the limitations of VASP in describing the electronic core-levels. The calculations were carried out with a spin-polarized fully relativistic approach (including the spin-orbit coupling). For the exchange-correlation potential, GGA with PBE parameterization was used. The fully localized limit of the GGA+U functional (actually, local spin density approximation – LSDA+U) [14], with the Hubbard U repulsion set to 0, 2 or 4 eV for Fe 3d orbitals, was used. An antiparallel configuration of magnetic moments on the two (top and bottom) Fe layers was assumed. The k-meshes equaled to 20×20×4 and the energy convergence criterion was set to $2.72 \times 10^{-5}$ eV ($10^{-6}$ Hartree). From the calculations, the density of states (DOS) of Fe 3p orbitals, the magnetic moments on iron atoms and the Mulliken charges on Fe and O atoms were determined.

**Additional data**

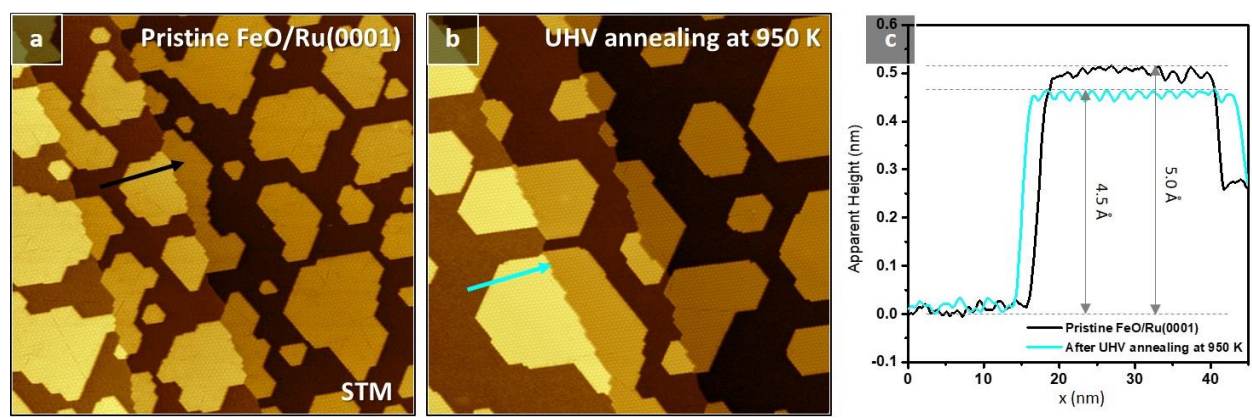

**Figure S1.** STM images of (a) pristine FeO/Ru(0001) and (b) the system after annealing at 950 K in UHV for 10 min. (c) shows height profiles drawn over iron oxide islands in (a) and (b) (the colors of the lines correspond to the colors of arrows in STM images). STM: $V_{sample}$=+1.0 V; $I_t$=0.7 nA; Images size: 200×200 nm$^2$.

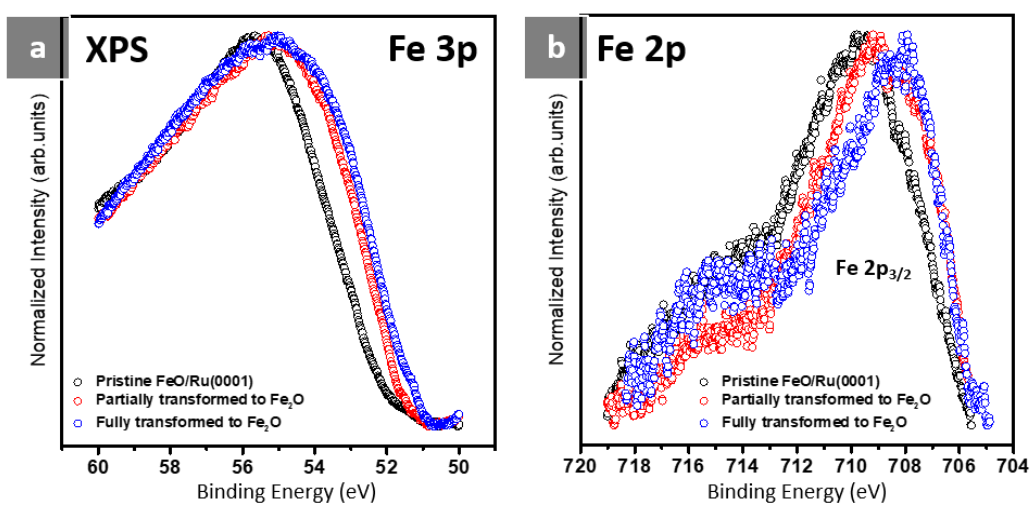

**Figure S2.** XPS Fe 3p (a) and Fe 2p (b) spectra obtained of pristine FeO/Ru(0001) (black symbols), as well as the system after brief (red) and prolonged (blue) annealing at 1050 K in UHV.

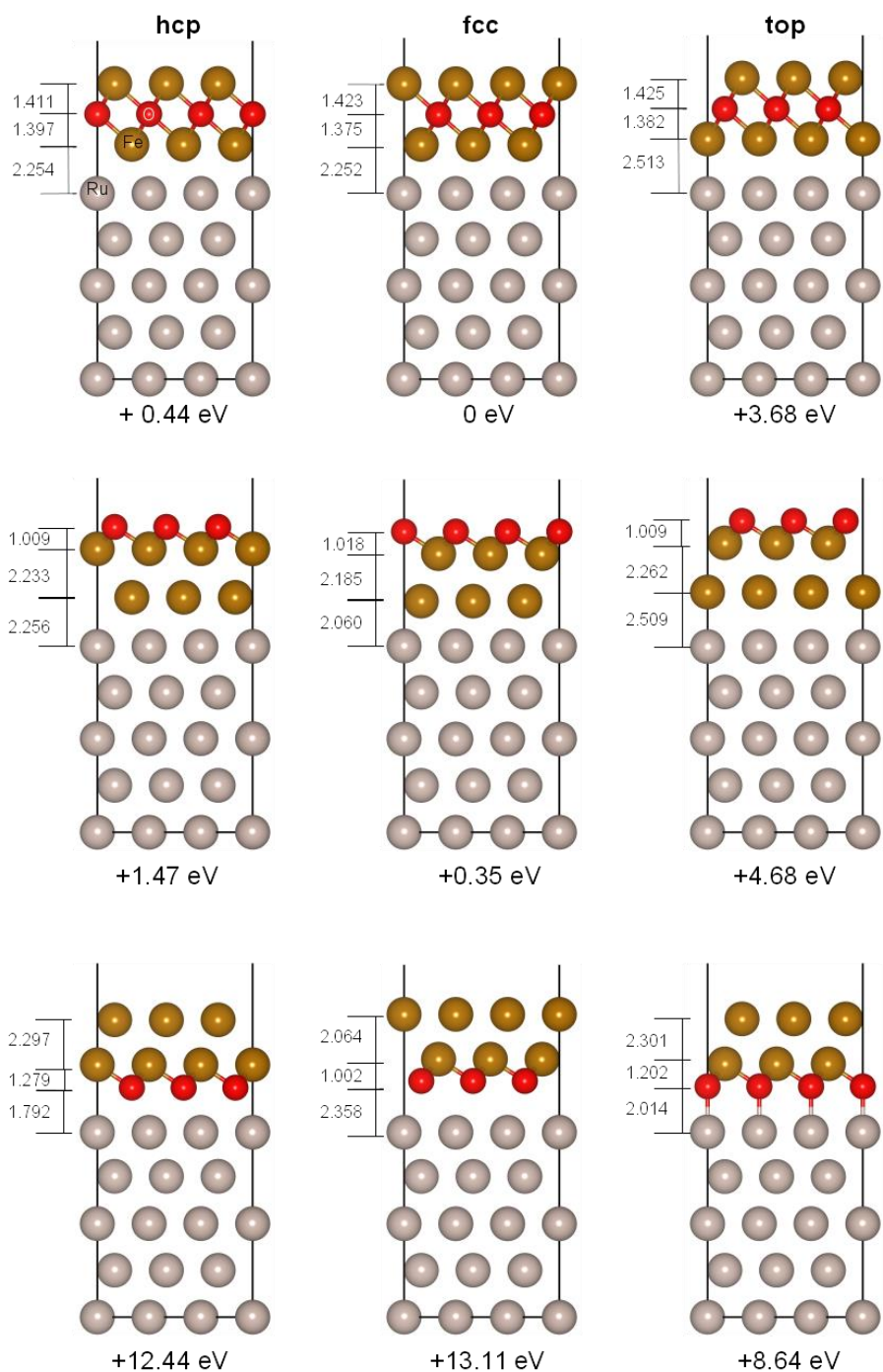

**Figure S3**. Calculated structures of differently stacked iron oxide films with Fe₂O stoichiometry adsorbed at different high-symmetry sites (*top*, *fcc*, *hcp*) of the Ru(0001) substrate. The calculations were performed for the iron oxide adopting the in-plane lattice parameter of the Ru substrate. Gold spheres represent Fe atoms, red O atoms and silver Ru atoms. All the layers were found to be uncorrugated after the relaxation. The total energy values are referred to the most stable system (*fcc* Fe-O-Fe/Ru). The distances are given in Å. The layer heights range from 5.05 to 5.78 Å. The lowest heights are observed for systems with Fe-O-Fe/Ru stacking, which is the natural stacking of FeO. For O-Fe-Fe/Ru and Fe-Fe-O/Ru layers arrangement, the layer heights are larger by more than 0.2 Å. Analysis of the stability of Fe₂O layers on Ru(0001) shows that the system with oxygen atoms at the interface is energetically

unfavored. This configuration can be described as two layers of Fe adsorbed on an O-precovered Ru(0001). The total energy of such configuration is by few eV higher than the most stable one. The Fe-O-Fe/Ru and O-Fe-Fe/Ru stackings are very close in total energy values. The latter can be considered as a monolayer of O adsorbed on two monolayers of Fe on Ru(0001).

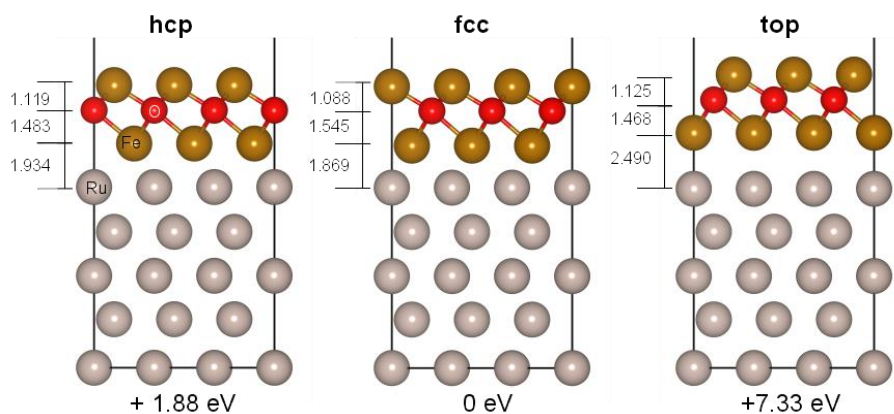

**Figure S4**. Calculated structures of Fe-O-Fe stacks adsorbed at different high-symmetry sites (*top*, *fcc*, *hcp*) of the Ru(0001) substrate. The calculations were performed for the Ru in-plane lattice parameter expanded to the one of iron monoxide (FeO). Gold spheres represent Fe atoms, red O atoms and silver Ru atoms. All the layers were found to be uncorrugated after the relaxation. The total energy values are referred to the most stable system (*fcc*). The distances are given in Å. The layer heights obtained within the *fcc* and *hcp* regions are roughly 0.5 Å lower than those obtained from calculations in which the iron oxide was adopting the in-plane lattice constant of the Ru substrate (compare with Figure S3), while the one calculated for the *top* region is approx. 0.2 Å lower.

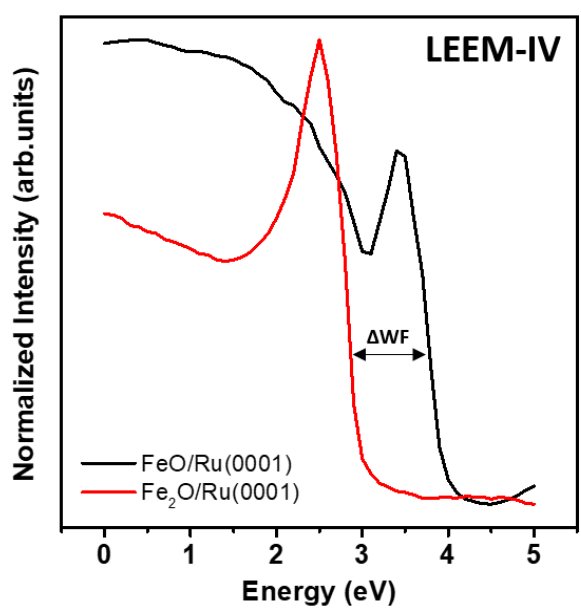

**Figure S5**. LEEM-IV spectra obtained for FeO/Ru(0001) (black) and $Fe_2O$/Ru(0001) (red).

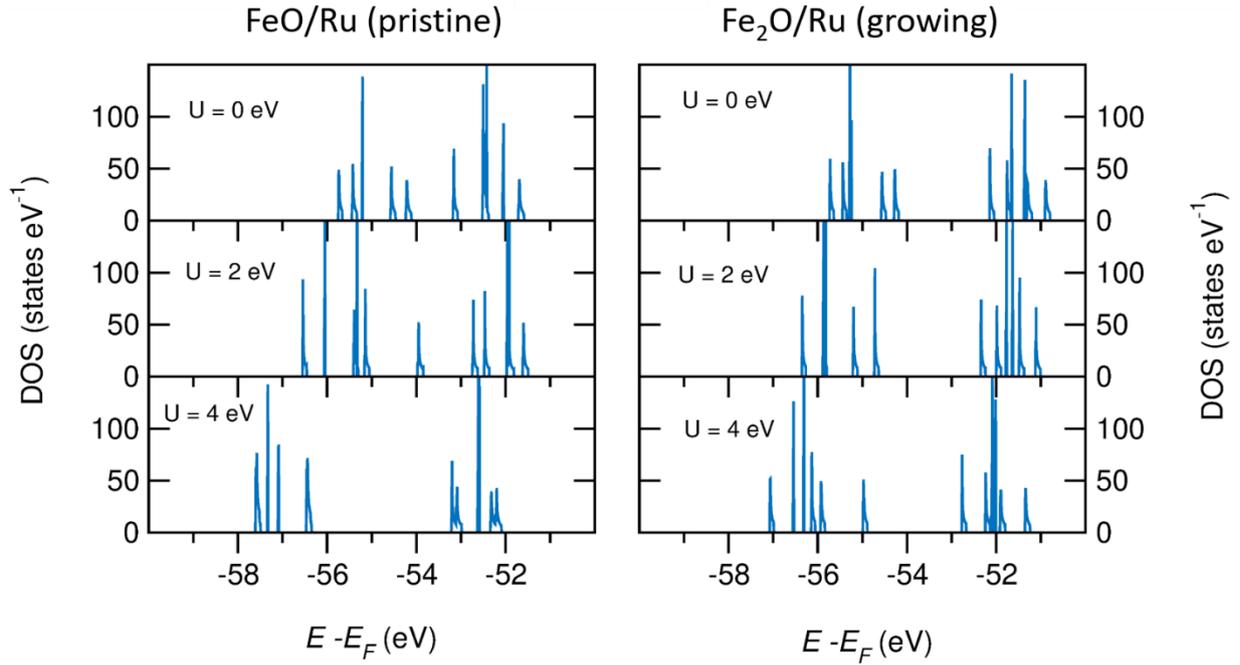

**Figure S6.** DOS of Fe 3p orbitals as a function of on-site repulsion $U_{3d}$ calculated for bilayer FeO/Ru and $Fe_2O$/Ru structures.

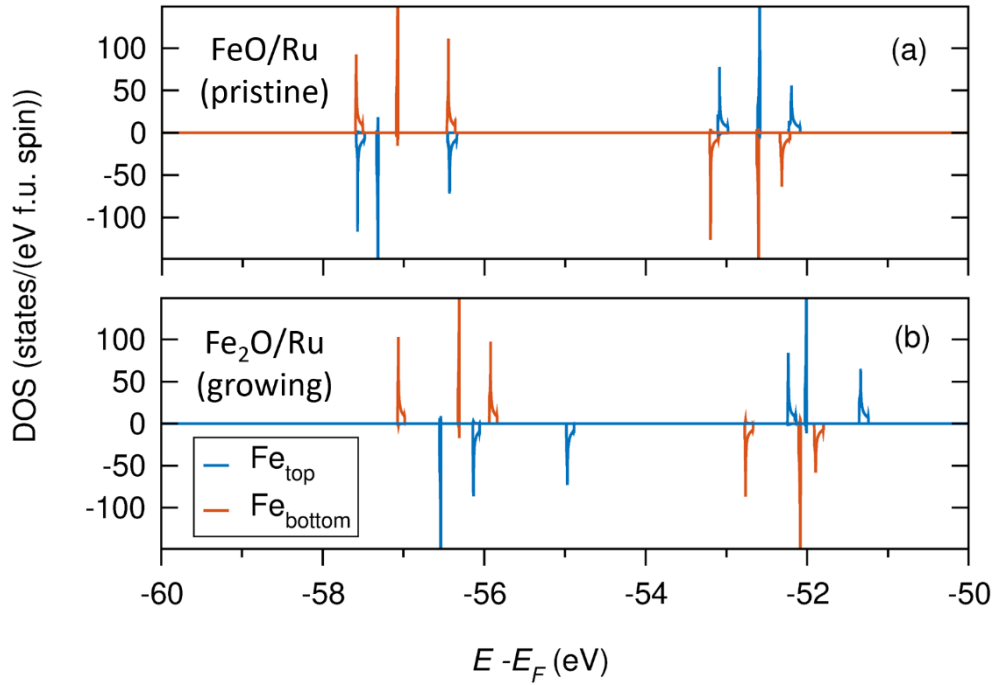

**Figure S7.** Spin-polarized DOS of Fe 3p orbitals calculated for (a) bilayer FeO/Ru and (b) $Fe_2O$/Ru structures.

**Table S1.** Calculated work function values of $Fe_2O$ films with different layers stacking adsorbed on Ru(0001).

| | Work function (eV) | | |
|---|---|---|---|
| Stacking | Fe-O-Fe/Ru | Fe-Fe-O/Ru | O-Fe-Fe/Ru |
| *fcc* | 4.65 | 4.89 | 7.65 |
| *hcp* | 4.47 | 4.35 | 7.67 |
| *top* | 4.64 | 4.46 | 7.77 |

**Table S2.** Spin ($m_s$) and orbital ($m_l$) magnetic moments, as well as Mulliken atomic charges calculated for bilayer FeO/Ru and $Fe_2O$/Ru structures with an antiparallel configuration of magnetic moments on the two (top and bottom) Fe layers. The calculations were carried out using the FPLO code in a fully relativistic approach, with PBE+U (U(Fe 3d) = 4 eV) and for the quantization axis [001].

| | FeO/Ru (pristine) | | | $Fe_2O$/Ru (growing) | | |
|---|---|---|---|---|---|---|
| | $m_s$ ($\mu_B$) | $m_l$ ($\mu_B$) | charge | $m_s$ ($\mu_B$) | $m_l$ ($\mu_B$) | Charge |
| Fe bottom | 4.00 | 0.09 | +0.58 | 3.77 | 0.07 | +0.46 |
| O middle | 0.05 | 0.00 | -0.87 | 0.04 | 0.00 | -0.84 |
| Fe top | -4.21 | -0.06 | +0.96 | -3.55 | -0.20 | +0.39 |
| O top | -0.31 | 0.00 | -0.63 | | | |